\documentclass[conference,letterpaper]{IEEEtran}
\usepackage[font=small,skip=10pt]{caption}
\usepackage{enumitem}
\usepackage{tikz}
\usepackage{color}
\usepackage{enumitem}
\usepackage{graphicx}
\usepackage{array}
\usepackage{float}
\usetikzlibrary{calc,trees,positioning,arrows,chains,shapes.geometric, decorations.pathreplacing,decorations.pathmorphing,shapes, matrix,shapes.symbols}
\newcommand{\comment}[1]{}

\tikzset{
>=stealth',
  punktchain/.style={
    rectangle, 
    rounded corners, 
    draw=black, very thick,
    text width=10em, 
    minimum height=2.5em, 
    text centered, 
    on chain},
  punktchain2/.style={
    rectangle, 
    rounded corners, 
    draw=none, very thick,
    text width=10em, 
    minimum height=2.5em, 
    text centered, 
    on chain},
  line/.style={draw, thick, <-},
  element/.style={
    tape,
    top color=white,
    bottom color=blue!50!black!60!,
    minimum width=8em,
    draw=blue!40!black!90, very thick,
    text width=10em, 
    minimum height=3.5em, 
    text centered, 
    on chain},
  every join/.style={->, thick,shorten >=1pt},
  decoration={brace},
  tuborg/.style={decorate},
  tubnode/.style={midway, right=2pt},
}

\begin{document}

%
\title{Identification of Bugs and Vulnerabilities in TLS Implementation for Windows Operating System Using State Machine Learning}

\author{
\IEEEauthorblockN{Tarun Yadav}
\IEEEauthorblockA{Scientist, Scientific Analysis Group\\Defence R \& D Organisation, INDIA\\Email: tarunyadav@sag.drdo.in}

\and

\IEEEauthorblockN{Koustav Sadhukhan}
\IEEEauthorblockA{Scientist, Defence Research and\\Development Organisation, INDIA\\Email: koustavsadhukhan@hqr.drdo.in}}

\maketitle

\begin{abstract}
TLS protocol is an essential part of secure Internet communication. In past, many attacks have been identified on the protocol. Most of these attacks are due to flaws in protocol implementation. The flaws are due to improper design and implementation of program logic by programmers. One of the widely used implementation of TLS is SChannel which is used in Windows operating system since its inception. We have used ``protocol state fuzzing'' to identify vulnerable and undesired state transitions in the state machine of the protocol for various versions of SChannel. The client as well as server components have been analyzed thoroughly using this technique and various flaws have been discovered in the implementation. Exploitation of these flaws under specific circumstances may lead to serious attacks which could disrupt secure communication. In this paper, we analyze state machine models of TLS protocol implementation of SChannel library and describe weaknesses and design flaws in these models, found using protocol state fuzzing.    
\end{abstract}

\IEEEpeerreviewmaketitle

\section{Introduction}

Transport Layer Security is the protocol responsible for secure communication over Internet. HTTPS, SFTP, SMTP and many other application layer protocols use TLS for secure communication. The protocol uses various cryptographic schemes like Asymmetric Key Encryption, Symmetric Key Encryption and Hashing to ensure confidentiality, authenticity and integrity of data. Vast use of TLS makes it a good target for security researchers and attackers. In the past, many attacks have been developed by the attackers which raised questions on security provided by the protocol, but with time the protocol has improved a lot. Most of these attacks target implementations of the protocol rather than the protocol itself. There are many implementations of TLS which are implemented by various programmers with their own understanding of the protocol. Many times individual understanding of the protocol specification do not cover all possible combinations of inputs and outputs, which leaves the protocol implementation vulnerable to attacks. To identify such vulnerabilities security researchers use many techniques. One of such technique is fuzzing which is widely use to find vulnerabilities in software implementations. In this paper we have used a technique called “protocol state fuzzing” which is used to identify undesired protocol state transitions in a specific protocol implementation. This technique is very useful in finding invalid inputs to a state, which may lead to a valid state with invalid transitions.  

The paper is organized into 8 sections. Section \ref{relatedwork}  describes the related work in this domain. Section \ref{overview} gives an overview of TLS protocol which explains handshake mechanism of the protocol. Section \ref{learning} discusses about learning procedure, SChannel implementation and experimental setup. Section \ref{models} explains design of state machine models, types of bugs and vulnerabilities and attack scenario. Section \ref{analysis} presents  learned models for various operating system and discusses analysis of these models. Section \ref{implications} explain implications of flaws found in the learned models and the paper ends with concluding remarks in section \ref{summary}.

\section{Related Work} \label{relatedwork}
There have been many analysis of TLS protocol which revealed various vulnerabilities in the protocol. Most of such analysis are focused on use of weak parameters or vulnerability in software implementation. DROWN\cite{sect2:1}, FREAK\cite{sect2:2}, LOGJAM\cite{sect2:3}, SLOTH\cite{sect2:4} are example of attacks which exploited  weak parameters in the protocol while HEARTBLEED\cite{sect2:5} is an example of vulnerability in software implementation.\\
Another dimension of TLS analysis is verification methods. There are mentions of verification methods which are used to verify correctness of protocol implementation with respect to protocol specification, but most of such literature provides abstract description rather than a practical implementation. Generally, verification methodology has an advantage that it not only verifies the correct path, but also finds the incorrect one. Such incorrect paths can be analyzed further to identify scope of attacks on protocol implementations. Such type of analysis is called "Protocol State Fuzzing", which provides all possible kinds of inputs to each state and traces the corresponding outputs. There are methods which apply such techniques to find bugs and vulnerability in the implementations. One such tool\cite{sect2:6} has been designed using machine learning and tested against various open-source implementations of TLS. In this paper we have extended this work further and analyzed Windows TLS library called SChannel which is a closed source implementation. 

\section{Overview of TLS Protocol} \label{overview}
\begin{figure}[!htb]
\centering
\includegraphics[width=8cm,height=12cm,keepaspectratio]{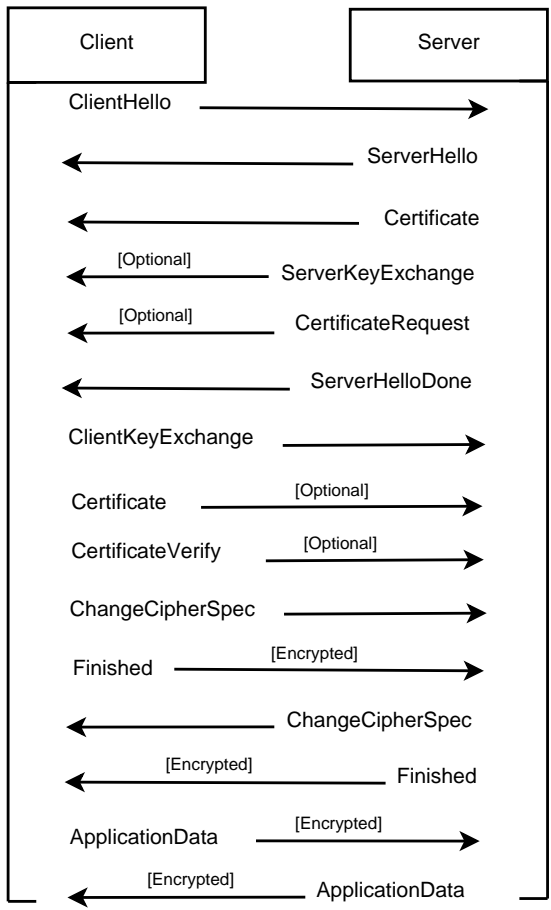}
\caption{TLS Handshake: Messages Exchanged between Client and Server}
\label{TLSHandshake}
\end{figure}

TLS is successor of SSL which was developed in 1994. SSL 2.0\cite{sect3:1} was the first public version of the protocol, which was deprecated very soon and improved to SSL 3.0\cite{sect3:2} which was the first stable version and supported many legacy systems. Later TLS 1.0\cite{sect3:3}, 1.1]\cite{sect3:4} and 1.2\cite{sect3:5} succeeded version SSL 3.0 and formally defined by IETF. \\
TLS Protocol consists of 2 layers, handshake and record layer. Handshake is responsible for negotiation of various algorithm and parameters while record layer pack and unpack every message. The messages which are used by handshake and record layer are specified and defined in respective RFCs by IETF. Fig.\ref{TLSHandshake} describes various messages exchanged during a handshake. 

As shown in Fig. \ref{TLSHandshake} messages are exchanged between a client and server to start a secure and encrypted session. In this process some of the messages are optional which are exchanged, only if systems are configured to use specific parameters. The following description briefly explains role and control flow of these messages while establishing a secure session using TLS protocol. 
\begin{enumerate}
\item{\textit{ClientHello:} The session is always initiated by the client with a ClientHello Message. It is the first message of the session and contains basic information for negotiation. This message specifies maximum version of the protocol supported, list of supported cipher suites (key exchange algorithm, encryption algorithm and hash function) and a random number which is used for key generation. Depending on the configuration, this message could have extensions too.}
\item{\textit{ServerHello:} It is a reply message from the server in response to ClientHello Message. It is the first message from server which specifies TLS version and cipher suite to be used for secure session. This message contains a random number similar to ClientHello message and it is used in key generation algorithm.}
\item{\textit{Certificate:} This message is sent from the server with a certificate owned by it. Client uses this certificate to authenticate the communicating server. RSA public key which is embedded in certificate is also used either in key exchange mechanism or for integrity check.}
\item{\textit{ServerKeyExchange[Optional]:} This message is mostly used when Diffie-Hellman key exchange is used. This contains various parameter needed from server side to generate a common secret for both client and server.}
\item{\textit{CertificateRequest[Optional]:} This message is uncommon and only used when server is configured to establish connection only with authenticated client. This message asks client to send a valid certificate to authenticate itself. }
\item{\textit{ServerHelloDone:} This is simple message which inform the client that as per protocol specifications all required information has been sent from the server .}
\item{\textit{ClientKeyExchange:} This message is sent from client to server with all necessary information to generate encryption keys. In case of RSA key exchange, a secret encrypted with RSA public key is sent to the server.}
\item{\textit{Certificate[Optional]:} This message contain a certificate owned by client and sent to the server to authenticate the client.This message is a reply to CertificateRequest message sent by the server.}
\item{CertificateVerify[Optional]:This message is sent to provide explicit verification of the certificate by signing the handshake messages sent and received by the client.  This message is only sent for the client certificate that has signing capability.}
\item{\textit{ChangeCipherSpec:} This message is one byte message and it is used to indicate the receiving side that messages following this message are encrypted with encryption key which is generated using information shared previously. }
\item{\textit{Finished: }This is first encrypted message sent from client to server. It is used to inform that handshake is finished and it also checks integrity of handshake messages to prevent session hijacking from MiTM(Man-In-the-Middle). This message contains hash of all handshake messages sent from sender of this message. The receiving side match this hash with the hash of handshake messages received from sender.}
\item{\textit{ApplicationData:} This message contains actual data of user, encrypted with encryption key from sender which is decrypted with same key on receiving side. Encryption and decryption keys are generated on both sides using information shared in handshake messages.}
\end{enumerate}

\section{Learning of TLS Protocol State Machine} \label{learning}
State Machine of a system indicates the system behavior for every kind of inputs to the system. Although it is expected that system must be designed as per specifications, but this does not always happen. While developing a system all possible cases are not considered, which are later analyzed and then exploited by attackers. \\
TLS protocol is a system of messages which are used to establish a secure connection between two users in a network. There are many kinds of messages which are exchanged during the process of TLS handshake. Upon receiving and processing a message the TLS system(client or server) outputs a message (including empty response) and changes its state to receive next input. State-Machine of a TLS system is behavioral representation of responses of the system for various kinds of inputs. Each state of state-machine is designated for specific kind of task which is established by receiving and responding particular kinds of input and output pairs. On completing the designated task, the state transfers the control to the next state, depending upon the current input and output pairs.\\
Protocol State Fuzzing uses the technique of fuzzing in which all kinds of inputs are given to each state and output responses are analyzed to form state-machine diagram of protocol. These responses are analyzed automatically using machine learning technique\cite{sect2:6}\cite{sect4:1}.
SChannel\cite{sect4:2} is a library which implements SSL/TLS protocol in Windows Operating System and it is stored in the system as schannel.dll. This library provides all the necessary functionalities for connection establishment, encryption and decryption of messages. This is not a open-source library so it works as black box without providing details of implementation. Most of the Microsoft services and third-party software for Windows use this library for secure Internet connection, therefore analysis of this library becomes more critical as it may affect a large number of systems around the world. 
Earlier development of learning system analyzed mainly open source libraries of TLS protocol and results were discussed\cite{sect2:6}. In this paper, we have improved the earlier development by implementing ECDHE key exchange mechanism in addition of RSA and DHE. We have also made changes to the system to run for all kinds of Windows OS. Other than these changes TLS clients have been developed for each version of Windows Operating System which were used to obtain client state machines of TLS implementations. Virtual machines have been used to learn the models and the model have been verified using FlexTLS library which provides direct access to individual message of the protocol. Design of such state machine models, and the types of bugs and vulnerabilities found using this approach are discussed in next section.

\section{State Machine Models} \label{models}
In this section we will discuss about design of learned models of TLS library of Windows operating system. These models represents a visual representation of how state transitions happen and where it may go wrong. In next subsection, we will discuss design of these models which will provide better understanding the approach used in this paper.

\subsection{Design of State Machine Models}
\begin{figure*}
\includegraphics[width=\textwidth]{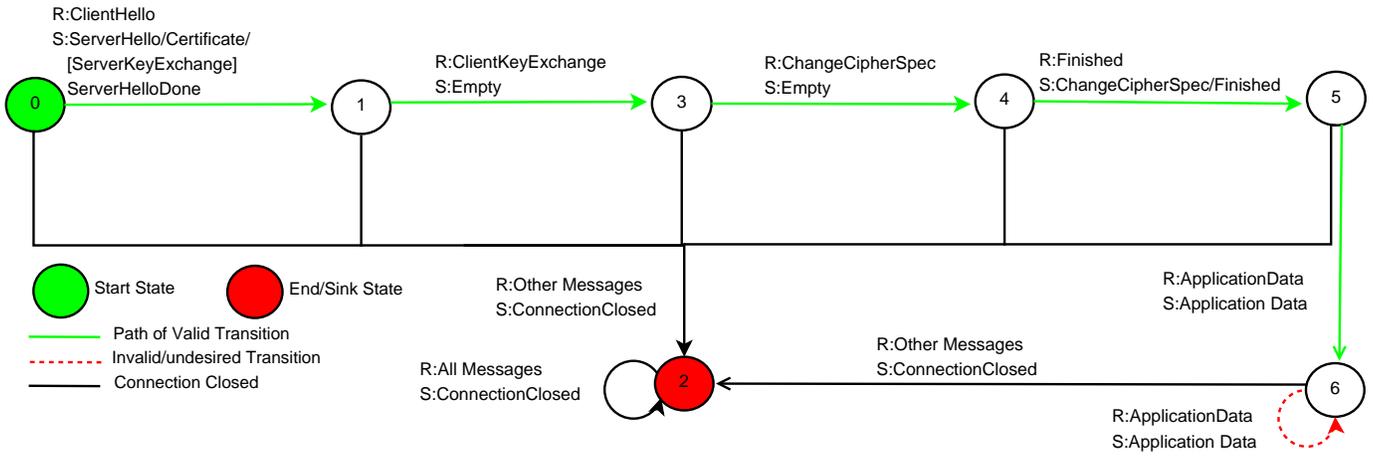}
\caption{State Machine Diagram of an Ideal TLS Implementation}
\label{IdealStateMachine}
\end{figure*}
Models described in this paper are based on a typical state machine diagram where each state is labeled with number and arrows are used to specify the transition from one state to another. Every model (e.g. fig. \ref{IdealStateMachine}) start with a green colored state labeled 0 and ends with red colored state which is labeled 2. In the context of TLS protocol, end(or sink) state is the state which closes the connection or absorbs all the inputs. Solid green arrows define a path of valid handshake while dotted red arrows indicate invalid or undesired transitions. Black arrows indicate transitions which terminate the ongoing connection. 

In figure \ref{IdealStateMachine} green path ($0\rightarrow1\rightarrow3\rightarrow4\rightarrow5\rightarrow6$) defines the valid handshake and for all the other messages connection is closed and control is passed to the end state, which is state 2. Figure \ref{IdealStateMachine} is the expected and ideal model of every TLS protocol implementation. In the next section, we will discuss practical behavior the state machines models which deviate from the ideal model. Due to deviation from expected model, bugs and vulnerabilities may arise in the system. Next subsection describes these bugs and vulnerabilities and discusses how these can effect the security of system and protocol.

\subsection{Types of Bugs and Vulnerabilities}
Undesired transitions in a state machine are indication of bugs which may lead to critical vulnerabilities in the system. These vulnerabilities can be further exploited by the attacker to compromise security and privacy of the communication\cite{sect5:1}\cite{sect5:2}. In context of TLS protocol, following types of bugs can exist in a state machine model, which are described in next section.

\begin{enumerate}
\item{\textit{Additional end(or sink) state}:  End/sink state is the state which absorbs every input and doesn’t change the state. In TLS state machine model, there must be only one such state which corresponds to closing of ongoing connection. Additional sink state hangs the system because it just processes the given input, but doesn't pass the control to any state. Such type of bugs may lead to denial-of-service vulnerabilities, where the sink state exhaust the CPU resources of the system to process the repeated crafted inputs fed by the attacker.  }
\item{\textit{Self-Loops}: Self-loops are the transitions in a state machine model which on a particular input remain in the same state. Self-loops are not expected in a state machine model unless it is specified explicitly. Similar to the sink state, existence of self-loops may lead to Denial of Service attacks where self loop exhausts CPU resources of the system by processing numerous crafted inputs provided by the attacker.}
\item{\textit{Alternate Paths}: Alternate paths in a state machine model are the paths from a valid state to another valid state using invalid states or transitions. Existence of alternate paths may lead to critical vulnerability, where it is possible to reach to an important state, bypassing necessary transitions from previous states and thus violating protocol specifications. Such vulnerability may become even more critical, if it bypasses the key exchange or authentication mechanism of TLS protocol. CCS injection\cite{sect5:3} is one of the such vulnerability which exploits an alternate path vulnerability and start encryption of messages without mandatory key exchange mechanism.}
\item{\textit{Undesired Replies}: Undesired replies are indication of poor implementation where states are not checked against every possible input messages. In such cases, connection is closed for particular input but before closing the connection an unnecessary or undesired reply is sent. Although such replies do not affect the security of system, but they might introduce abnormal behavior in the system which violates the TLS protocol specifications.}
\end{enumerate}

\subsection{Attack Scenario}
In this paper we have mentioned about attacks which can cause various kinds of damages to the system and the ongoing communication. By mentioning attacks we mean two types of attack scenarios. First type of attack scenario is Denial-of-Service where attacker acts as a client (or server) and establish many connections to its counterpart server (or client). Due to existing vulnerabilities these connections are accepted and processed further, which could result in heavy consumption of system resources and system may deny services provided by it.\\ Another type of attack scenario is Man-In-The-Middle(MiTM) where attacker comes in between of client and server and impersonate it as an other end of communication (client or server). Using this technique attacker can modify order and content of protocol messages and due to vulnerabilities described in next section, such crafted messages are accepted by the system which could weaken the security of communication. 

\section{Analysis of SChannel Based State Machine Models} \label{analysis}
\begin{figure*}[!h]
\includegraphics[width=\textwidth]{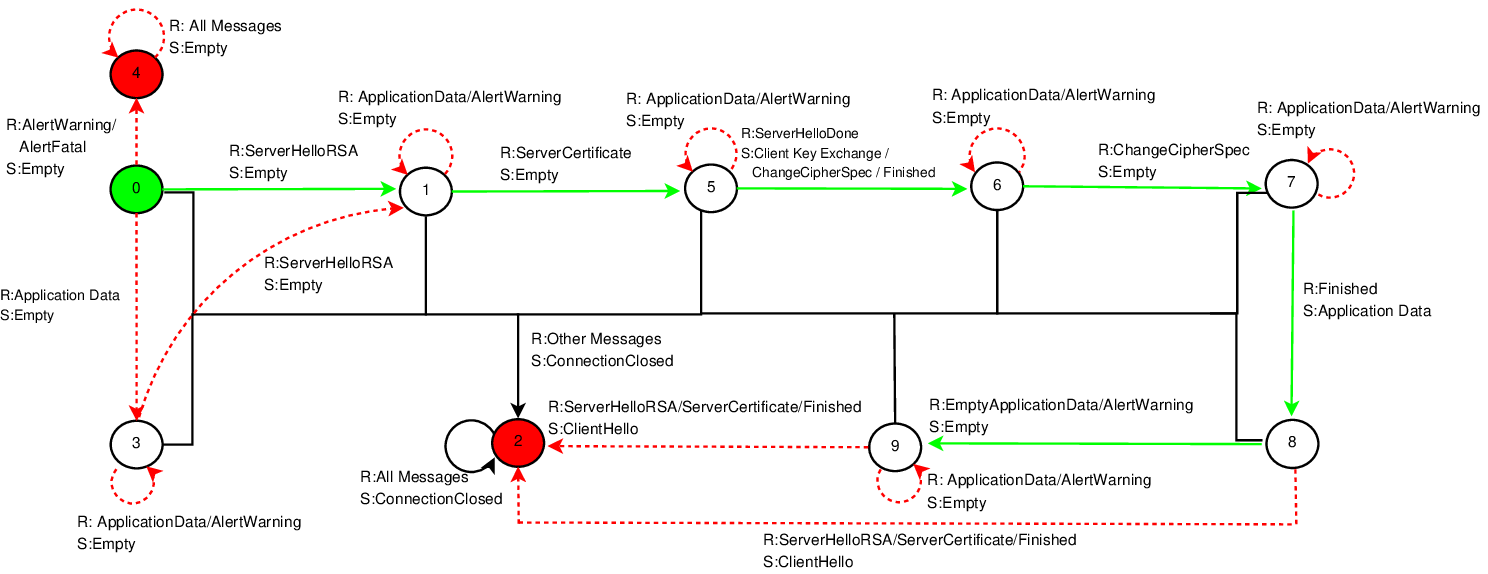}
\caption{OS: Windows 7; TLS Version: 1.0; Key Exchange: RSA (Client Implementation)}
\label{windows7}
\end{figure*}

\begin{figure*}[!h]
\includegraphics[width=\textwidth]{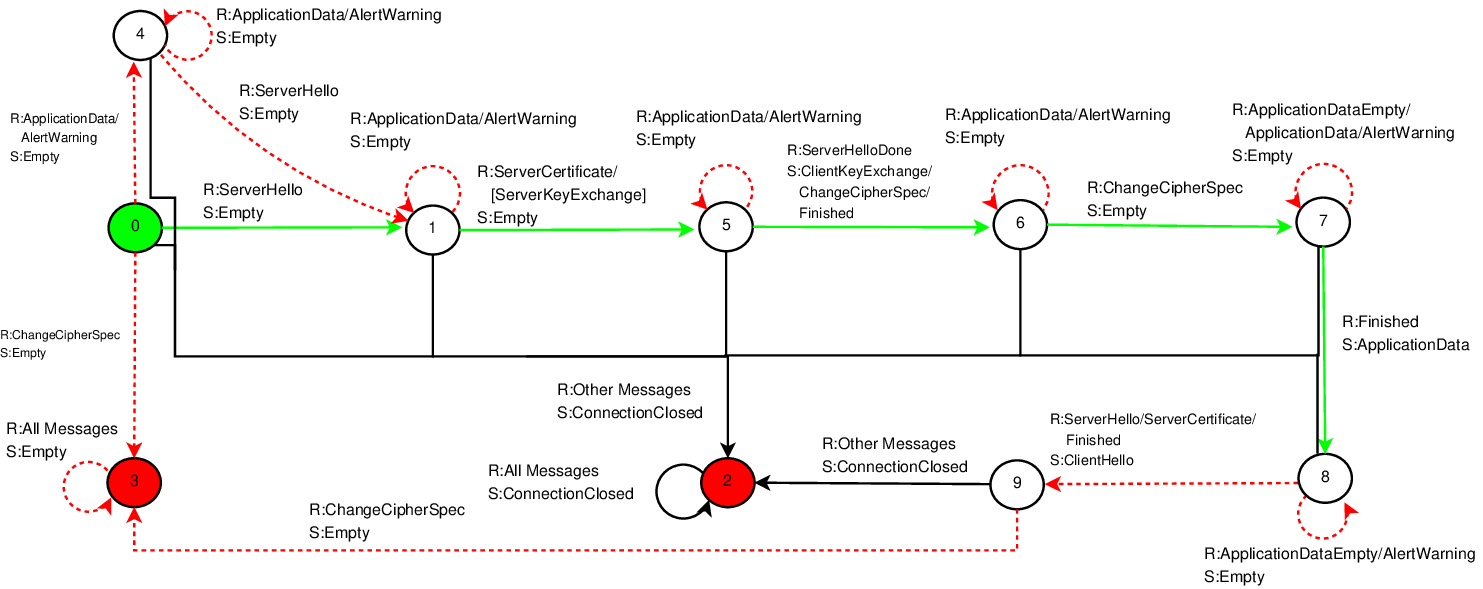}
\caption{OS: Windows 8 and 10; TLS Version: 1.2; Key Exchange: RSA/ECDHE (Client Implementation)}
\label{windows810}
\end{figure*}

We have designed and analyzed various state machine models for different combinations of Windows and TLS versions. Most of these models have implementation bugs except server model for Windows Server 2016. From the state machine models, it is clear that with newer versions implementation has been improved and number of undesired states has been reduced. Simple structure and less no. of states indicates a better and secure version of state machine model. In this section, we analyze state machine models of Windows SChannel Library for various version and discuss about bugs and vulnerabilities found in the system. 
\begin{table}[h!]
\begin{center}
\begin{tabular}{ | m{.6cm} | m{4cm} | m{1cm}| m{1.7cm} | } 
\hline
\bf{Figure No.} & \bf{State Machine Model}& \bf{Sink States} & \bf{States with Self Loops} \\ 
\hline
3 & Windows 7 RSA TLS 1.0 (Client) & 2,4 & 1,4,5,6,7,9 \\ 
\hline
4 & Windows 8 RSA TLS 1.0 (Client) & 2 & 0,1,2,3,4,5,6 \\ 
\hline
5 & Windows 8 and 10 RSA TLS 1.2 (Client) & 2,3 & 1,3,4,5,6,7,8 \\ 
\hline
6 & Windows Server 2008 RSA TLS 1.0 (Server) & 2,3 & 1,3,4,5,6 \\ 
\hline
7 & Windows Server 2012 (Server)& 2 & 1,4,5,6,7 \\
\hline
8 & Windows Server 2016 (Server) & 2 & 1,3,4,5 \\
\hline
\end{tabular}
\label{analysis_table}
\caption{Sink States and Self Loops in State Machine Models}
\end{center}
\end{table}
As we have discussed in last section that more than 1 sink state and self loops are the bugs which may lead to denial of service kind of vulnerabilities. Table-1 describes sink states and self-loops for each state machine model. Figure no. \ref{windows7}, \ref{windows810} and \ref{server2008} have extra sink states which may lead to denial of service attack, where these state absorbs every input without passing the control flow. Every model has some states with self loop which allows  to absorb the inputs without changing the state which may again lead to critical denial of service attacks, if the connection does not close after a timeout period. Following subsections will discuss each model in detail.
\subsection{Windows 7 TLS 1.0 RSA \\\ [Client Implementation] (Figure \ref{windows7})}
\begin{figure*}[!h]
\includegraphics[width=\textwidth]{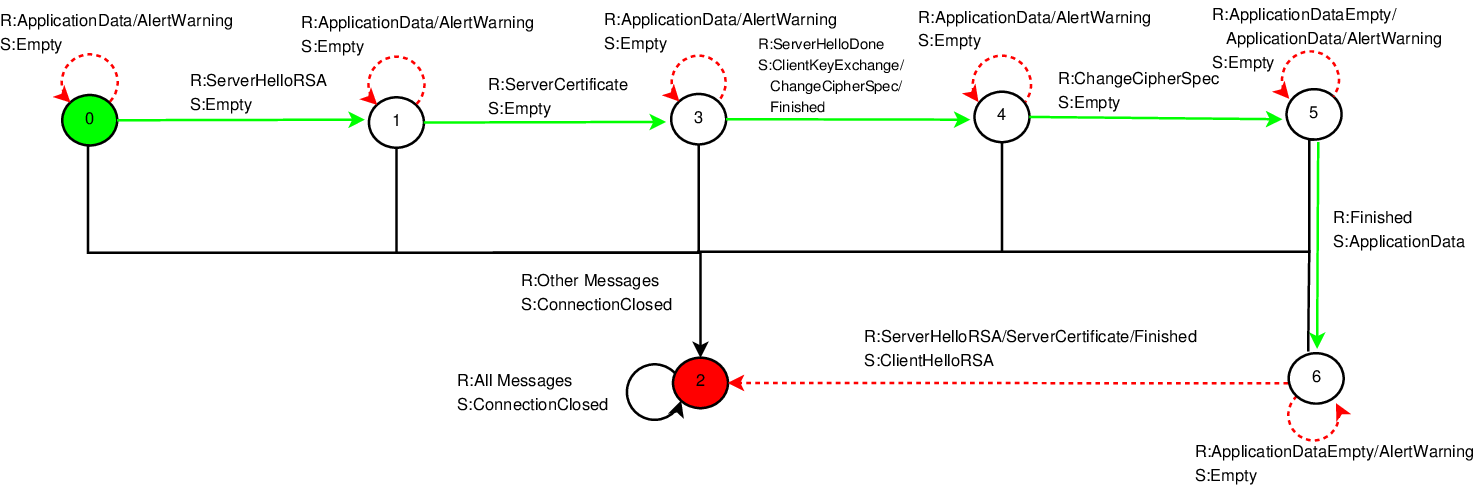}
\caption{OS: Windows 8; TLS Version: 1.0; Key Exchange: RSA (Client Implementation)}
\label{windows8}
\end{figure*}
\subsubsection{Valid Transitions (Handshake)}
\begin{description}
\item[]
\quad \space \space A valid handshake exists in the figure \ref{windows7} which is shown using green arrows (path $0\rightarrow1\rightarrow5\rightarrow6\rightarrow7\rightarrow8\rightarrow9$). This path validates the transitions mentioned in discussed figure \ref{IdealStateMachine}  which specifies expected behavior of protocol for a TLS client implementation.
\end{description}

\subsubsection{Invalid/undesired Transitions}

\begin{itemize}
\item{In figure \ref{windows7} there is an extra sink state 4, which is created by supplying Alert messages to the start state 0. This sink state absorbs all the inputs and disrupts the connection by owning the control flow forever. Other than additional sink state, states 1, 5, 6, 7, 9 have a self-loop for two inputs called ApplicationData and AlertWarning. These self-loops signify that for these mentioned two inputs, the state doesn't change and it absorbs the inputs without any output. In this case too, states own the control for a particular message instead of closing the connection.}
\item{Alternate path to reach state 1: In figure \ref{windows7} transitions through state 3 create an alternate path to reach state 1. State 0 receives ApplicationData message and makes a transition to state 3 which on receiving ServerHelloRSA message changes the state to state 1. Existence of state 3 shows misunderstanding of programmer while implementing the protocol. Any input message for state 0, except ServerHelloRSA, must close the connection such that it should not lead to an alternate path to some valid state. Existence of alternate path may skip states of valid handshake which may lead to a handshake with invalid parameters. CCS Injection is one of such vulnerability which exploits existence of alternate path to a valid state.}
\item{Undesired replies to messages in state 8 and 9:  For three kinds of input messages types ServerHelloRSA, ServerCertificate and Finished abnormal behavior is detected in state 8 and 9 where instead of closing the connection ClientHello is sent and state is changed to 2. Although this ClientHello doesn't initiate a new connection because after reaching to state 2, next input message closes the connection. These transitions again shows weak implementation, where states are not checked against every possible input messages.}
\end{itemize}

\subsection{Windows 8 TLS 1.0 RSA \\\ [Client Implementation] (Figure \ref{windows8})}

\subsubsection{Valid Transition (Handshake)}
\begin{description}
\item[]
\quad \space \space 
Similar to figure \ref{windows7}, figure \ref{windows8} also depicts a valid handshake for Windows 8 client implementation which is shown using green arrows (path $0\rightarrow1\rightarrow3\rightarrow4\rightarrow5\rightarrow6$). This path validates the transitions mentioned in discussed figure \ref{IdealStateMachine} which signifies the existence of a successful handshake.
\end{description}
\subsubsection{Invalid/undesired Transitions}
\begin{itemize}
\item{As we can see states 0, 1, 3, 4, 5 have a self-loop for two inputs called ApplicationData and AlertWarning. State 5 has a self-loop for another input called ApplicationDataEmpty. State 6 has a self-loop for ApplicationDataEmpty and AlertWarning message. As mention in section \ref{models}, these loops may exhaust CPU resource if multiple connections are started frequently. }
\item{Undesired replies to messages in state 6:  For three kinds of input messages types ServerHelloRSA, ServerCertificate and Finished abnormal behavior is detected in state 6 where instead of closing the connection, ClientHello is sent and state is changed to 2. Although this ClientHello doesn't initiate a new connection because after reaching to state 2, next input message terminates the connection. Existence of such state machine bugs show lack of understanding of protocol specification specified in RFCs.}
\end{itemize}

\subsection{Windows 8 and 10 TLS 1.2 RSA \\\ [Client Implementation] (Figure \ref{windows810})}
\subsubsection{Valid Transition (Handshake)}
\begin{description}
\item[]
\quad \space \space 
In figure \ref{windows810} a valid handshake is shown using green arrows (path $0\rightarrow1\rightarrow5\rightarrow6\rightarrow7\rightarrow8$). Existence of this path validates the handshake transitions required for TLS 1.2. This handshake is validated for RSA and ECDHE key exchange mechanism and both resulted in similar state transitions for a successful handshake.
\end{description}

\subsubsection{Invalid/undesired Transitions}
\begin{itemize}
\item{In figure \ref{windows810}, state 3 is an extra sink state which is created by receiving ChangeCipherSpec message at the start of connection. This state can also be reached from state 9 with same input. Once control is reached to state 3, it is never passed to any other state for any input and therefore sink state 3 is created. There are states (1, 3, 4, 5, 6, 7, 8) which have self-loops for mainly two inputs called ApplicationData and AlertWarning. These self-loop with no outputs result in idle transitions which only absorb given inputs and don't play any meaningful role in the system of states.}
\item{Alternate path to reach state 1:  Upon receiving ApplicationData or AlertWarning messages at the start of connection instead of ServerHello, the connection doesn't close and control goes to state 4 which creates an alternative path to the state 1. This alternate path is similar to the transition we have discussed in figure \ref{windows7}. State 4 passes the control to state 1 upon receiving ServerHello message and this behavior is similar to state 0. As this path doesn't bypass any authentication or key exchange message it is not a security vulnerability but an implementation bug which deviates from the path of valid handshake. }
\item{Undesired replies to messages in state 8 and 9: Similar to state 9 of figure \ref{windows7}, state 8 replies with ClientHello message and passes the control to state 9 upon receiving ServerHello or ServerCertficate or Finished message. In this case, expected behavior was to close the connection but ClientHello message is sent to the server. This ClientHelloRSA doesn't initiate a connection because from state 9 all connection are closed for all inputs except ChangeCipherSpec. Therefore this bug doesn't qualify for a security vulnerability. State 9 transfers the control to state 3 with no output upon receiving ChangeCipherSpec message but after state 3 every input is absorbed and system doesn't output any message. Although these undesired replies and wrong transitions are not security issues, but must be fixed to prevent exploitation of such bugs in future.}
\end{itemize}

\begin{figure*}
\includegraphics[width=\textwidth]{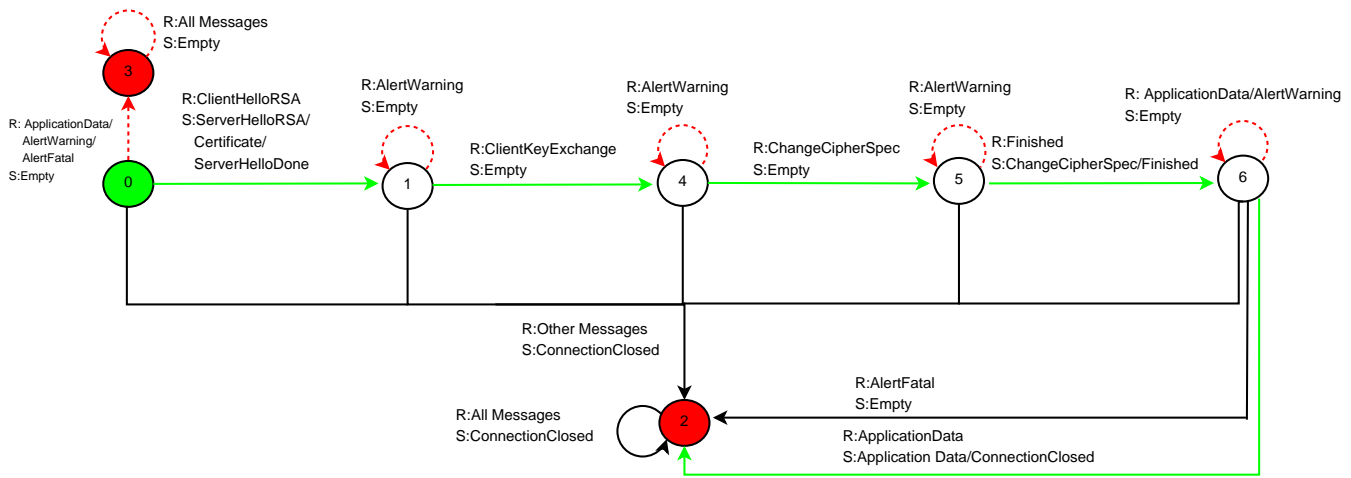}
\caption{OS: Windows Server 2008; TLS Version: 1.0; Key Exchange: RSA (Server Implementation)}
\label{server2008}
\end{figure*}

\subsection{Windows Server 2008 TLS 1.0 RSA  \\\ [Server Implementation] (Figure \ref{server2008} )}
\begin{figure*}
\includegraphics[width=\textwidth]{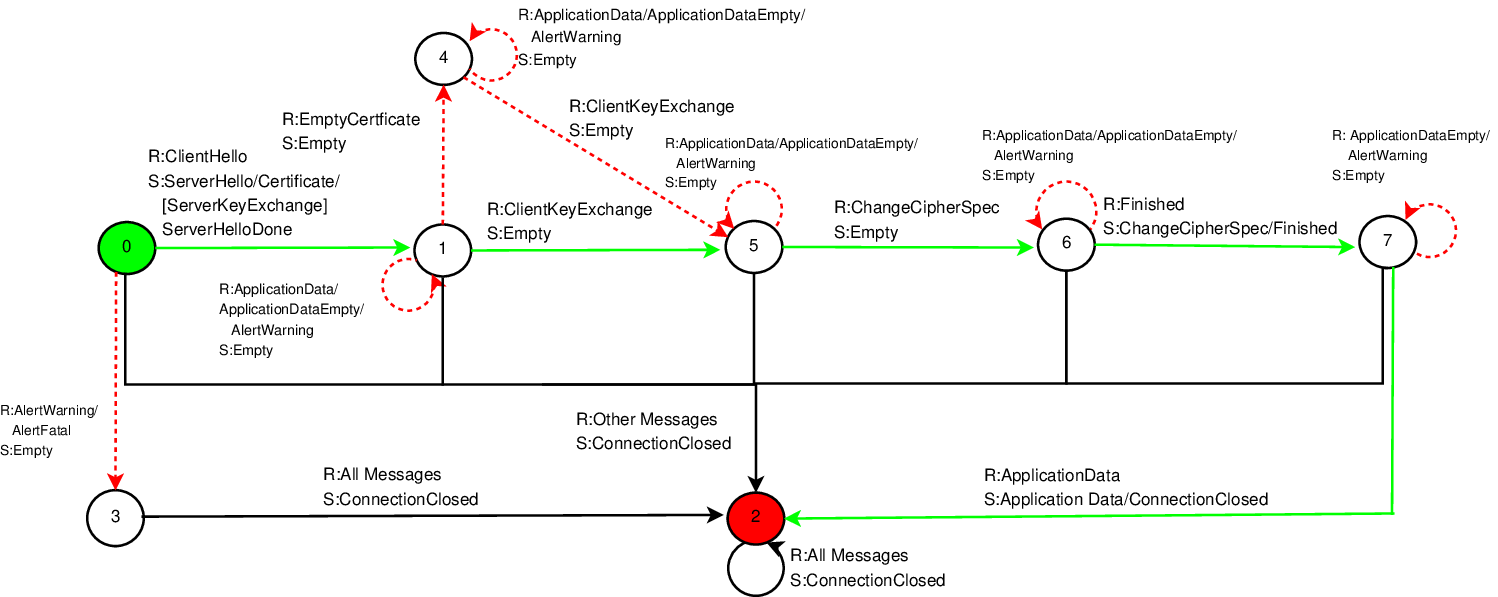}
\caption{OS: Windows Server 2012; TLS Version: 1.0-1.2; Key Exchange: RSA/ECDHE (Server Implementation)}
\label{server2012}
\end{figure*}
\begin{figure*}
\includegraphics[width=\textwidth]{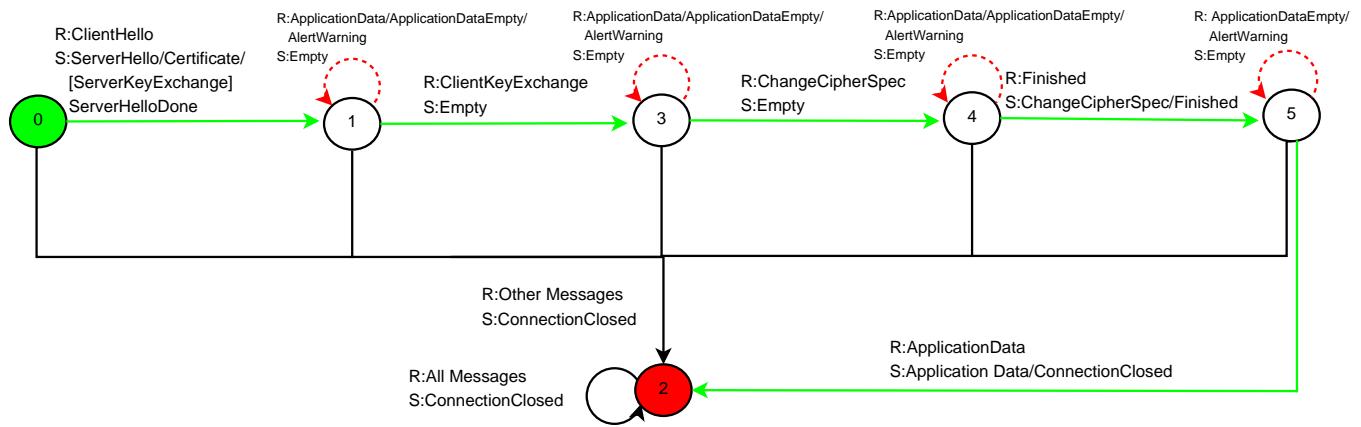}
\caption{OS: Windows Server 2016; TLS Version: 1.0-1.2; Key Exchange: RSA/ECDHE (Server Implementation)}
\label{server2016}
\end{figure*}

\subsubsection{Valid Transition (Handshake)}
\begin{description}
\item[]
\quad \space \space 
In figure \ref{server2008} a valid handshake is shown using green arrows (path $0\rightarrow1\rightarrow4\rightarrow5\rightarrow6\rightarrow2$). This path validates the transitions mentioned in figure \ref{IdealStateMachine} for a valid handshake. 
\end{description}
\subsubsection{Invalid/undesired Transitions}
\begin{itemize}
\item{Similar to model described in figure \ref{windows7} for Windows 7, a sink state is created when receiving ApplicationData or Alert message as the first message to communication. This sink state accepts all message with empty reply and loops the control flow of connection to itself. States 1, 4, 5, 6 have a self-loop each for two inputs called Alert-Warning and ApplicatonData(only for state 6). Upon receiving these inputs, mentioned states don't respond and wait for next input without changing the state. These states receive such inputs till the connection timeout and then closed forcefully.}
\end{itemize}

\subsection{Windows Server 2012 \\\ [Server Implementation] (Figure \ref{server2012})}

\subsubsection{Valid Transition (Handshake)}
\begin{description}
\item[]
\quad \space \space 
Similar to state machine model for Windows Server 2008 green arrows (path $0\rightarrow1\rightarrow5\rightarrow6\rightarrow7\rightarrow2$) in figure \ref{server2012} depicts state transitions for a valid handshake. This path is observed for not only for RSA but also for ECDHE key exchange mechanism and in both the cases similar state machine model is learned. 
\end{description}
\subsubsection{Invalid/undesired Transitions}
\begin{itemize}
\item { In figure \ref{server2012} states 1, 4, 5, 6, 7 have a self-loop for two inputs called ApplicationData and AlertWarning. These self-loops signify that for these mentioned two inputs state is not changed and the state absorb the inputs without responding any output.}

\item{Alternate path to states 2 and 5: In figure \ref{server2012} transitions through state 4 create an alternate path to reach state 5 from state 1. There is one more transition through state 3 to reach state 2 from state 0. State 0 receives AlertWarning/AlertFatal message and make a transition to state 3 which on receiving any input message changes the state to state 2. Both transitions are unexpected but doesn't meet the requirement of security vulnerability as no security parameter is bypassed or changed using these transitions. Similar kind of behavior is observed in figure \ref{windows810} where existence of alternate paths is described. This behavior shows that similar bugs which existed in client implementation, also exists in server implementation. Reuse of client code in server implementation could be a possible reason for such improper design. Therefore designers and programmers must understand and design control flow of protocol messages separately for client and server.  }
\end{itemize}

\subsection{Windows Server 2016 \\\ [Server Implementation] (Figure \ref{server2016})}

\subsubsection{Valid Transition (Handshake)}
\begin{description}
\item[]
\quad \space \space 
Figure \ref{server2016} describes a valid handshake which is shown using green arrows (path $0\rightarrow1\rightarrow3\rightarrow4\rightarrow5\rightarrow2$). This path is almost similar to the one which is shown in figure \ref{IdealStateMachine} and have lesser number of undesired states than others. By comparing this model with other models it is quite clear that this model is more correct and simpler than previous state machine models. Most of the times simpler models are less vulnerable due to lesser complexity of state transitions which leaves lesser area for vulnerable attack surface. It is also clear that latest versions of SChannel are continuously improving by approaching towards ideal state machine model.
\end{description}
\subsubsection{Invalid/undesired Transitions}
\begin{itemize}
\item{In figure \ref{server2016} states 1, 3, 4, 5 have a self-loop for three types of inputs  ApplicationData, ApplicationDataEmpty and Alert-Warning. Although this model is simpler and better but self-loops are still there which may put the system in unnecessary computation and wait condition, instead of closing the connection. }
\end{itemize}

\section{Implications} \label{implications}
In this paper, we have discussed about various kinds of bugs and vulnerabilities present in state machines of various versions of Windows TLS library called SChannel. Windows operating system is one of the most popular operating system, therefore presence of a bug in Windows  directly affects many computer systems around the globe. We have discussed in previous sections, how these bugs can cause denial of service attacks, which can shutdown a critical Windows based server of an organization to create panic to the users and financial loss to the organization. \\
To protect the system from such attacks, it is always necessary for developers as well as for testers to verify each possible transition of a protocol implementation. Leaving one loophole in implementation may cause severe vulnerability in the system. Understanding the protocol specification is also an important part of protocol implementation and it must be done as a group task so that individual understanding of the specification doesn't influence the implementation.\\

\section{Summary} \label{summary}
This paper presents state machine models for TLS implementation of Windows operating system. These models have been designed using query based machine learning technique. The paper describe various bugs and vulnerabilities present in these model due to incorrect implementation of TLS protocol in SChannel library. In this paper, we have analyzed a subset of messages that are exchanged between a client and a server but we have not included state machine behavior for many optional messages and various extensions of messages. Analysis of complete set of these messages could reveal more bugs in the implementation which could cause severe attacks on secure communication.\\

\end{document}